**Book chapter:**

**Fostering learners' self-regulation and collaboration skills and strategies for mobile language learning beyond the classroom**

Authors:

**Olga Viberg** (KTH Royal Institute of Technology, Sweden). Email: oviberg@kth.se

**Agnes Kukulska-Hulme** (The Open University, UK). Email: agnes.kukulska-hulme@open.ac.uk

**Abstract**
Many language learners need to be supported in acquiring a second or foreign language quickly and effectively across learning environments beyond the classroom. The chapter argues that support should focus on the development of two vital learning skills, namely being able to self-regulate and to collaborate effectively in the learning process. We base our argumentation on the theoretical lenses of self-regulated learning (SRL) and collaborative learning in the context of mobile situated learning that can take place in a variety of settings. The chapter examines a sample of selected empirical studies within the field of mobile-assisted language learning with a twofold aim. Firstly, the studies are analyzed in order to understand the role of learner self-regulation and collaboration while acquiring a new language beyond the classroom. Secondly, we aim to provide a deeper understanding of any mechanisms provided to develop or support language learners' self-regulated and collaborative learning skills. Finally, we propose that fostering SRL and collaborative learning skills and strategies will benefit from recent advances in the fields of learning analytics and artificial intelligence, coupled with the use of mobile technologies and self-monitoring mechanisms. The ultimate aim is to enable the provision of individual adaptive learning paths to facilitate language learning beyond the classroom.

**Introduction**
Many second language learners, including older K-12 students, university students and migrant adult learners, are in great need of guidance and support in developing sufficient and appropriate language skills to participate in higher levels of education or to get a job. Language classes alone do not provide sufficient opportunities to progress quickly and in ways that are tailored to the specific needs of people in these groups. Moreover, in view of the recent widespread moves to remote online teaching and learning worldwide (Czerniewicz et al.,

2020), effective learner support in settings beyond the physical classroom has become even more critical (Viberg, Wasson, & Kukulska-Hulme, 2020).

Researchers have highlighted that learners in increasingly online learning settings can benefit from the enactment of self-regulated and collaborative learning strategies, skills and knowledge. Self-regulated learning (SRL) can influence positively and predict learners' academic performance (Zimmerman, 1990; Viberg, Khalil, & Baars, 2020) and their ability to acquire a second language efficiently (Oxford, 2016; Botero, Questier, & Zhu, 2019; Viberg & Andersson, 2019; Viberg, Khalil, & Bergman, 2019; Yang, 2020). Collaborative learning activities, assisted by the use of mobile technologies, can "create opportunities for practising language skills and building new knowledge and relationships inside and outside the classroom, as well as in settings where there are no classrooms but there may be other meeting spaces or joint activities" (Kukulska-Hulme & Viberg, 2018, p. 208). Hence, in this chapter we argue that support for language learners should: (1) focus on fostering language learners' *self-regulated learning* and *collaborative learning* strategies and skills that will enable them to take better control over their own learning processes across learning environments, and (2) carefully consider the design and situated use of mobile technologies, which have been found to be beneficial for second language learners in developing those strategies and skills (e.g., Shadiev, Liu, & Hwang, 2019; Viberg, Mavroudi, & Ma, 2020; Zhang, Cheng, & Chen, 2020). We also argue that in the mobile-assisted language learning design process, we need to build on recent advances in the fields of learning analytics (see e.g., Viberg, Wasson, & Kukulska-Hulme, 2020) and artificial intelligence (e.g., Dodigovic, 2020) to be able to offer adaptive language learning paths that address the learners' needs across learning settings.

To examine these issues in more detail, we identified and analysed a sample of published empirical studies concerned with learner self-regulation and collaboration when learning a new language in and beyond the classroom. The sample was bounded by limiting the publication period to peer reviewed papers and chapters in the most recent decade (2010-20) and searching with keywords related to mobile devices and learning (e.g., smartphones, iPads, mobile assisted language learning, situated language learning, contextual language learning). Related search terms for 'self-regulated language learning' included self-directed, self-regulated and autonomous learning, while related search terms for 'collaborative language learning' included cooperative, group and social learning. In these studies, learning was sometimes a combination of in-class and out-of-class (or outdoor) activities, but we excluded papers that were entirely focused on in-class learning. When making a final selection for our analysis (total of 20 papers), we chose studies where there was a clear concern with the development of skills and strategies and/or consideration of technology or learning designs. In the subsequent analysis of the chosen papers, we paid particular attention to any mechanisms provided by researchers, teachers or system/learning designers to develop or support language learners' self-regulated and collaborative language learning. The results of this analysis contribute to the key constructs and key issues presented in the sections below.

When presenting the key constructs, i.e., self-regulation in language learning and collaborative language learning in relation to mobile contexts, we start with an introduction to the terms, followed by presentation of: 1. central underlying theoretical concepts, 2. the relation of the constructs to mobile learning (m-learning) and mobile-assisted language learning (MALL), 3. how they have been examined in the setting of out-of-class MALL activities, 4. what specifically has been studied, and finally, 5. what support mechanisms have been offered.

**Key constructs**

**Self-regulation in (mobile) language learning**

Self-regulated learning (SRL) refers to "the process by which learners personally activate and sustain cognitions, affects and behaviors that are systematically oriented toward the attainment of learning goals" (Zimmerman & Schunk, 2011, p.vii). Strategic SRL is also central and critical to second language acquisition (Oxford, 2011). 'Strategic' points to the way/s in which learners approach learning tasks – offered by the teacher, the intelligent tutor, or the learner if the task is self-initiated – by choosing and enacting a range of relevant strategies and tactics that they believe are "best suited to the situation, and applying those tactics appropriately" (Winne & Perry, 2000, p. 533). Self-regulated language learning involves the use of specific *metastrategies*, *strategies* and *tactics* (Oxford, 2011) that can empower learners to take control of their own learning, and hence to acquire the target language more effectively, for example in terms of time spent and effort expended on achieving language learning goals.

Language learners can use self-regulated language learning strategies to regulate several interrelated aspects of their learning, including their beliefs, behaviours, their internal mental states and the learning environment (Oxford, 2011). The enactment of these strategies is especially critical for being able to regulate their out-of-class language learning activities, assisted by mobile technologies (Botero et al., 2019). This can be explained by the fact that language learning in out-of-class settings can be undermined by other priorities and obligations in learners' lives. As demonstrated in research findings, learners are not using MALL out-of-class as often as one would expect (e.g., Dashtestani, 2016), and overall "students tend to not self-direct using technology" (Botero et al., 2019, p. 73).

Despite the fact that SRL is beneficial for learner success, it is challenging for learners, because many have difficulty calibrating their own learning processes (Stone, 2000; Viberg, Khalil, & Baars, 2020) and they have poor SRL skills and knowledge when planning, monitoring and reflecting on their learning activities (e.g., Baars & Viberg, *in press*; Bjork, Dunslosky, & Kornell, 2013). Scholars also stress that learners are not often aware of effective learning strategies and how to employ them (e.g., Cervin-Ellqvist et al., 2020; Dirkx et al., 2019). Nevertheless, SRL strategies and skills can be learnt and taught (Lodge et al., 2018), and to assist learners, teachers and learning designers in this task, we can take advantage of the affordances offered by mobile technology-in-use that can provide relevant individual SRL support at the right place and the right time (e.g., Viberg, Mavroudi, & Ma, 2020).

SRL theoretical lenses are grounded in a number of SRL models (for overview, see Panadero, 2017) that can be used to underpin the design of relevant support mechanisms for self-regulated language learning. One of these models is Zimmerman's SRL model (Zimmerman, 2002). It explains the SRL cyclical process through three phases of self-regulation: the *forethought*, *performance* and *self-reflection* phases. In the *forethought* phase, learners plan their learning activities by, for instance, examining the task (or formulating it if the learning activity is self-initiated), setting their short- and/or long-term goals, and planning how to achieve them. In the *performance phase*, learners carry out the task by monitoring the learning progress and controlling targeted learning activities or actions through the use of selected self-controlling strategies and tactics that keep them engaged so that they complete the task. In the final *self-reflection phase,* learners evaluate their learning performance and reflect on it by considering reasons behind successes and failures. Based on Zimmerman's SRL model, researchers have offered theoretical models of language learner self-regulation. One of them is the *Strategic Self-Regulation* (S2R) model of language learning (Oxford, 2011, 2016). Hitherto, this model has been largely used to monitor and understand various aspects of language learners' self-regulation in their second language acquisition rather than to underpin the design of relevant support mechanisms that would enable the effective use of SRL metastrategies, strategies and tactics (Oxford, 2011) by the learner when acquiring a new language. Peeters et al. (2020) employed the S2R model to reveal second language learners' SRL tactics in an academic writing course. Köksal and Dundar (2016) used it to develop a scale for the use of self-regulated language learning strategies. Others (Saqr et al., *in press*) used the S2R model to code language learners' SRL tactics to uncover their dynamics in a computer-supported collaborative learning setting. Based on such understandings, effective ways of supporting language learners in their self-regulated language learning beyond the classroom should be developed and offered.

In a recent review examining the association between m-learning and SRL, Palalas and Wark (2020) reported that the results of the majority of the reviewed studies (*n*=38), including MALL papers, showed that m-learning enhanced learners' SRL and SRL enhanced m-learning, suggesting that mobile technology can be used effectively to foster students' SRL and the other way around, i.e., learners' ability to self-regulate their learning process is advantageous for m-learning and can improve learning outcomes.

In the context of MALL, some scholars have shown that SRL behaviour is the most critical factor in predicting linguistic outcomes (Tseng, Cheng, & Hsiao, 2019). Another study found that language learners' self-regulatory capacity can be facilitated through MALL practices and that self-regulatory capacity plays an important role in determining the magnitude of associations between learning satisfaction and learning intention in MALL settings (Hwang, 2014). Furthermore, researchers have shown that MALL positively affects the enactment of students' SRL behaviours. Kondo et al. (2012) showed that a MALL module encouraged students' self-study in terms of time spent on learning tasks, levels of satisfaction derived from the tasks, and self-measured achievement.

Recent research highlights the importance of *out-of-class contexts* to support learners' engagement with MALL. Botero et al. (2019) examined language learners' engagement with a MALL tool (Duolingo app) and concluded that most students need training and support for their self-directed out-of-class learning. Another study introduced and evaluated a virtual-reality game-based English m-learning application for language learners' SRL (Chen & Hsu, 2020). The results show that the tool influenced students' SRL. They also found that self-efficacy and self-regulation were positively related to each other, i.e., students who had higher confidence believed that they were competent; they were using cognitive strategies, and were increasingly self-regulating using metacognitive strategies. In sum, there is a positive interdependence between SRL and MALL.

**Collaborative (mobile) language learning**
Collaborative learning is an approach that recognises the value of social interaction in learning, both from an affective perspective (many people are motivated through learning with others) and from the point of view of learning effectiveness. It is generally derived from Vygotsky's (1978) sociocultural theory of cognitive development, whereby social, cultural and historical forces play a part in an individual's development, and learning is facilitated by social interactions, particularly a learner's interactions with a more knowledgeable person. A key idea is that of the Zone of Proximal Development, which is the gap between what an individual can do on their own and what they can only achieve with some assistance. Crucially, that assistance can come from another learner, not only a teacher. Nowadays, it can also come from a computer-based (increasingly mobile) system playing a similar role. Collaboration can also be seen as a way to promote joint knowledge construction through discussion, reasoning, elaboration and debate.

It has long been acknowledged that collaborative learning is hard to define since it can involve various numbers and configurations of people, the meaning of 'learning' is open to interpretation, and it can be "a truly joint effort" or the learners may divide up the work among themselves (Dillenbourg, 1999). There are also differing perspectives on whether learning together should result in a product/outcome, or if the process of learning is the main focus. Importantly, "the words 'collaborative learning' describe a situation in which particular forms of interaction among people are expected to occur, which would trigger learning mechanisms, but there is no guarantee that the expected interactions will actually occur" (Dillenbourg, 1999, p. 5). Effort therefore needs to be put into planning and supporting interactions among the learners. In cases where collaboration is supposed to take place without support from a teacher, learners can feel helpless; in one study, there was not much discussion among the learners as they felt confused and were "each waiting for someone else to provide a better understanding" of an article they had all read (Chang-Tik & Goh, 2020, p. 7).

In language learning, collaboration offers opportunities for 'collaborative dialogue' (Swain & Watanabe, 2012), for example one in which learners help each other solve language-related problems such as what to say and how to say it. It can also be an opportunity for learners to engage in more language practice: more speaking, listening, reading, writing, memorising, rehearsing, especially as a way to extend this activity beyond the classroom. Additionally, peer feedback and corrections enable learners to reflect on their language use and knowledge, which can in turn improve the quality of their language production. Collaborative learning also exposes learners to social, emotional, cultural and multilingual dimensions of interaction (Baker, Andriessen, & Järvelä, 2013; Kukulska-Hulme & Lee, 2020; Walker, 2018), often beyond what is available in their classroom.

Research studies reporting technology-supported collaborative activities in language learning tend to report on increased opportunities to communicate, and with more people. It is argued that use of internet-based social media offers opportunities for social networking that may promote language learning (Zourou & Lamy, 2013). MALL extends those opportunities even further through easy and frequent access and thanks to an abundance of applications that enable communication, image, video and audio capture and sharing in multiple settings. In published studies, perceived and documented benefits of mobile collaboration for language learning are centred on: increasing opportunities for communication and negotiation in the target language (Berns et al., 2016; Ilic, 2015) including through peer feedback, commenting, reviewing and rating (Chai, Wong, & Lind, 2016; Hoven & Palalas, 2013; Hwang et al., 2014); listening and speaking practice with reflection on quality (Hwang et al., 2016; Pellerin, 2014); language use during co-creation of artifacts (Hoven & Palalas, 2013; Fomani & Hedayayi, 2016); improving quality and quantity of writing (Hwang et al., 2014); and practising cognitive strategies such as paraphrasing and summarization (Hazaea & Alzubi, 2016).

In the papers reviewed for this chapter, specific collaboration skills and strategies are generally not reported, but there are indications that teachers orchestrate collaboration in various ways. Andujar (2016) describes how a collaborative mobile instant messaging activity to develop ESL writing was carefully structured by the teacher and that there was daily tracking of each student's involvement in the activity. Kirsch (2016) used storytelling to encourage collaboration. In Berns et al. (2016) and in Tai (2012) students had well defined roles in the collaborative tasks. Hoven & Palalas (2013) mention procedures that students had to follow; while Hwang et al. (2014) report that students were instructed on how to give 'meaningful' feedback to their peers. Fomani and Hedayayi (2016) claim that it would be beneficial to train students in artifact creation.

**Key issues**

Fostering learners' SRL and collaborative learning skills and knowledge for MALL beyond the classroom is a challenging and complex task. There are several interrelated challenges but also some promising results that suggest these approaches are worth pursuing.

*First*, there is complexity in the 'self-regulated learning' and 'collaborative learning' concepts. Since these concepts consist of many interrelated aspects, dimensions and characteristics, it becomes challenging to select what concrete learning activities and/or aspects of individuals' self-regulated language learning (e.g., goal setting strategies or self-reflection activities) and collaborative language learning processes (e.g., social or cognitive dimension of collaborative learning) should be further developed and adequately supported. This suggests that researchers or practitioners need to identify a problem to be examined and related learning activities to be further developed and supported. In our sample, scholars have frequently studied some chosen aspect/dimension of self-regulation or collaboration in MALL settings, with limited attention to the cyclical and multifaceted nature of these processes. Chen et al. (2018) presented an English vocabulary m-learning app with an SRL mechanism–with a focus on developing learners' goal-setting skills when learning new words–to improve learning performance and motivation. Results indicated that the learners who used the tool with the SRL support exhibited significantly better learning performance and motivation than those who used the app without the SRL support. Others focused on the affective aspect of SRL and showed that: 1. the use of mobile technologies in second language learning can decrease anxiety for both learners and teachers (Kim, 2018), and 2. a designed Affective Learning SRL app for supporting second language learners with affective learning in their SRL process was seen to increase their awareness of the self-regulated language learning process and their engagement in, and motivation for, self-regulated learning across settings (Viberg, Mavroudi, & Ma, 2020). Also, scholars have focused on fostering migrant learners' time management skills when acquiring a host language in out-of-class settings and found that using the TimeTracker app to keep track of the time spent on studying a host language enabled learners to devote more time to their second language learning and become more engaged in it through monitoring of their learning activities (via a learning dashboard) and raising awareness of their learning process (Viberg, Khalil, & Bergman, 2019). Andujar's study (2016), which focused on development of students' accuracy as well as lexical and syntactic complexity in their ESL writing, demonstrated that daily out-of-class collaborative interactions on WhatsApp on their mobile phones improved the students' accuracy.

*Second,* the definition of m-learning is problematic, and several different definitions of m-learning and MALL are found in the reviewed sample. As stated by Grant (2019), m-learning has become an "'umbrella term for the integration of mobile computing devices within teaching and learning" (p. 361), and the term has been used unsystematically. For this reason, he instead recommends the use of "design characteristics that are essential to mobile learning environments" (p. 368). They are described as follows: Learner is mobile; Device is Mobile; Data services are persistent; Content is mobile; Tutor is accessible; Learner is engaged; and Physical and networked cultures and contexts impact learning or learner (Grant, 2019). These characteristics should be carefully considered when designing relevant support mechanisms aiming to foster language learners' collaborative and SRL strategies and skills in out-of-class MALL settings.

*Third,* there are difficulties in establishing whether there are language learning gains arising from student collaboration and their ability to self-regulate their own learning processes. In several projects (e.g., Berns et al. 2016; Hoven & Palalas, 2013), learning activities had both individual and collaborative elements that were closely interlinked, so any learning gains cannot be directly attributed to collaboration. Similarly, in Hwang et al. (2014), benefits derived from the situated nature of the activity are intertwined with benefits of peer feedback during collaboration. This finding is supported by earlier research emphasizing that successful collaboration in computer-supported collaborative learning settings requires diverse types of support, including support for promoting individual self-regulatory skills and strategies, peer support, facilitation of self-regulatory competence within the group, and socially shared regulation of learning (Järverlä et al., 2015). This suggests that all these factors may be critical for improved language learning performance in collaborative mobile language learning settings. Also, it has been shown that overall, there is a lack of alignment between the purpose of the tools in supporting self-regulation in online learning settings and the evaluation performed to assess their effectiveness (Pérez-Álvarez, Maldonado-Mahauad, & Pérez-

Sanagustín, 2018). In the same review, the scholars found that most of the studies did not evaluate the effect on learners' SRL strategies. This finding relates also to the papers reviewed for this chapter. Similarly, collaboration skills and strategies for MALL are usually not explicitly assessed in the studies we reviewed.

*Fourth,* there is a challenge in terms of *how* selected aspects of language learners' self-regulation and collaboration should be effectively supported in regard to underlying theoretical lenses of SRL and collaborative learning. In our sample, researchers seldom explicitly present their selected theoretical grounds when designing and examining MALL activities aiming to support language learners' SRL, and this is a limitation since considerable related research is available. For collaborative mobile language learning, relevant theories or frameworks are often mentioned at the start of a paper (e.g., social constructivism) but it is not clear how exactly they were applied in the learning designs. However, there are some exceptions. Kondo et al. (2012) for example, introduced the MALL module based on Zimmerman's model of self-regulation (Schunk & Zimmerman, 1998; Cleary & Zimmerman, 2002). This module was designed to facilitate language students' SRL "in the class, elsewhere in the university, or outside the university" (Kondo et al., 2012, p.174). The same SRL model was used to underpin the design of the Affective Learning SRL mobile app to support Japanese language learners in their self-study of the targeted second language (Viberg, Mavroudi, & Ma, 2020). Botero et al. (2019) applied the lenses of self-directed learning (Garrison, 1990) to study self-directed language learning in a mobile, out-of-class context. To fill the existing gap in the SRL theoretically underpinned MALL designs and tools, researchers have recently offered a conceptual framework, Mobile-Assisted Language Learning for Self-regulated learning (MALLAS; Viberg, Wasson, & Kukulska-Hulme, 2020) aimed at learning designers and grounded in the theoretical lenses of Zimmerman's model (2002) and also the S2R model of language learning (Oxford, 2011, 2016). Hwang et al. (2014) based their research in the value and types of peer feedback within collaborative language learning and used Stanley's (1992) four-step procedure to improve and structure peer feedback.

*Moreover,* there are several technology-related challenges. These include how we can best design and benefit from existing and emerging technologies (e.g., artificial intelligence (AI), virtual reality, visualization technologies) in combination with mobile technology to adequately address the needs of learners as well as to support them in the development of their SRL and collaborative learning skills when acquiring a new language. Another issue is how to design sustainable adaptive MALL solutions that would support individual learners in their self-study across learning environments. These challenges entail a need to carefully consider learning and technology design (see e.g., Chen & Hsu, 2020; Viberg, Wasson, & Kukulska-Hulme, 2020), policy-related issues that can contribute to the sustainability of MALL designs over time and space (e.g., the implementation of bring-your-own-device policy, see e.g., Bartholomew, 2019; Chen & Hsu, 2020), and mobile data-driven language learning solutions that would facilitate the provision of adaptive language learning paths (Viberg, Wasson, & Kukulska-Hulme, 2020). Generally, as emphasized by Pérez-Paredes et al. (2019), the potential of data-driven learning in the MALL context is underexplored.

*Finally*, there are challenges pertaining to learners' individual characteristics, including their level of motivation for language learning, cultural orientedness, self-regulation capacity and learning orientations (e.g., integrative and instrumental). Regarding learners' self-regulation capacity, Tseng et al. (2019) state that little is known "about the effects of different types of learning orientations and implementation intentions upon the development of self-regulated capacity in L2 m-learning context" (p. 371). Furthermore, they examined the causal relationships between goal orientation, implementation strategies, and SRL behaviour in relation to MALL, and found that SRL behaviour is the most important factor in predicting linguistic outcomes.

**Implications**

The development of MALL tools aimed at assisting learners in their use of SRL and collaborative language learning strategies in out-of-class settings is a multifaceted task that requires several careful design considerations. Since SRL and collaborative learning strategies and skills can be not only learnt but also taught, such tools can be aimed at learners and teachers, for example in the form of learner- or teacher-centered learning dashboards that visualize different aspects of the learning process. In the development of such tools, we need to thoroughly consider what indicators of SRL and collaborative learning activities should be visualized (and how) to best address the stakeholders' needs. Knowing which indicators influence stakeholders' understanding of the learning process will lead to a more actionable learner behaviour, and providing extra support on the part of the teacher.

Teacher support and just-in-time feedback is critical, especially at early stages of self-regulation and collaboration. This suggests that in the design process of MALL apps or services, we need to consider *how* such feedback can be provided. It is important to not only consider existing solutions and technologies but also to involve the stakeholders during the design process to best meet learners' needs and preferences.

When developing data-driven MALL approaches, several aspects have to be thought of. First, we need to consider what specific learning activities are targeted and how they will be assessed to be able to track the development of learners' SRL and collaborative strategies and skills. Second, there is a need to carefully think of what learner data to collect and how to collect it, in relation to the development of relevant learning analytics modules that comprise data, analytics and action (for more, see Viberg, Wasson, & Kukulska-Hulme, 2020). Third, since the use of mobile devices allows access to learners' contextual data (e.g., in out-of-class settings), it is of utmost importance to address the issues of learners' privacy and information security. The protection of learner data "is not only a fundamental right among others but the most expressive of the contemporary human condition" (Rodotá, 2009, p.82). This is not only a legal obligation, but also a moral one. To enact a more responsible use of mobile technologies for language learning, the designers of MALL tools should address key data protection principles associated with lawfulness, fairness, and transparency; purpose limitation; data minimisation; accuracy; storage limitation, integrity and confidentiality (European Union, 2017, p. 17).

Designed support for SRL and collaboration in MALL should consider the earlier mentioned essential m-learning characteristics listed by Grant (2019), and how to deal with possible challenges. For example, when learners are mobile, their individual interactions with an application and other learners are likely to vary as to frequency, regularity and intensity. When physical and networked cultures and contexts impact learning, learners may have reduced capacity to pay attention to a task in environments where they are susceptible to distractions, and they may be influenced in their interaction styles by parallel casual conversations with their friends on social media on the same device. Technology designs as well as teaching and learning practices will need to evolve to take account of expected challenges and to capture emerging ones that we may not yet know about. The teacher/tutor should be accessible, but that is not always practical and needs to be carefully managed or supplemented by intelligent tools. Crucially, if teachers have not been trained in SRL and collaborative learning, they need to become familiar with these approaches before they can support their students.

**Future directions**

Overall, considering that the use of mobile devices is beneficial for fostering learners' self-regulated- and collaborative language strategies and the other way around (i.e. for effective use mobile technologies for second language acquisition in out-of-class contexts, learners are expected to self-regulate their learning process and collaborate effectively with their peers, teachers and the learning environment), and an increasing interest in learning analytics for SRL (Viberg, Khalil, & Baars, 2020) and for understanding and supporting collaborative learning activities (Saqr, 2018), future pathways to provide adequate MALL support should apply data-driven approaches to learning. Such approaches should be based on sound theoretical lenses and take into account recent advances in the fields of AI and learning analytics to provide more individualized and personalized language learning paths. This will enable learner agency in out-of-class settings. In language learning and teaching tasks, AI can be employed to emulate the behaviour of a teacher or a learner (Dodogovic, 2020). This is important since the teacher is often unavailable in such settings, and access to peers can be likewise limited.

In the development and evaluation of adaptive self-regulated and collaborative learning paths, we also recommend taking a responsible approach to the protection of learners' privacy, ethics and security (see, Viberg, Wasson, & Kukulska-Hulme, 2020). In this regard, there is a need for empirical studies that would target different stakeholders (learners, teachers, school and university leadership, learning designers and developers) and enable more responsible use of student data in out-of-class MALL activities. Another future direction of study is to further validate relevant existing models and frameworks (e.g., Tseng et al., 2019; Viberg, Wasson, & Kukulska-Hulme, 2020) in different cultural and educational contexts. We also need more studies that are explicitly conducted to support L2 learners in out-of-class settings.

**Reflection questions**

Based on the argumentation and findings presented in this chapter, we offer several questions for the stakeholders' reflection.

For *the teacher*: What SRL and collaboration skills and strategies should your students have, if they are to be effective in the language learning activities or tasks they are expected to undertake, and how might they be developed?

For the *learning designer*: What m-learning design characteristics should be considered in combination with recent advances in the fields of learning analytics and AI to provide adaptive language learning paths?

For *the researcher:* How can you assist learning designers and teachers in the development of relevant support mechanisms, the use of which would provide better conditions for L2 learners' self-regulated and/or collaborative learning in out-of-class settings?

**Recommended reading**

To deepen the reader's understanding of how learners' self-regulation and collaboration in second language acquisition can be developed through MALL, we suggest the following literature:

Grant, M. (2019). Difficulties in defining mobile learning: Analysis, design characteristics, and implications. *Educational Technology Research and Development, 67*(2), 361–388.

To better understand existing challenges in defining m-learning, we recommend this paper which reviews definitions of m-learning and argues that they are not helpful in guiding the design of m-learning environments. Instead, it offers a framework of design characteristics for m-learning environments. It also presents implications for future research and instructional design.

Viberg, O., Wasson, B., & Kukulska-Hulme, A. (2020). Mobile-assisted language learning through learning analytics for self-regulated learning (MALLAS): A conceptual framework. *Australasian Journal of Educational Technology, 36*(6), 34-52.

To better grasp how mobile self-regulated language learning can be enacted through learning analytics, we suggest this paper which offers a theoretically grounded conceptual framework and guidelines to its operationalization. The paper includes practice guidelines for teachers in terms of task design and learner support.

Pishtari, F., Rodriguez-Triana, M., Sarmiento-Marques, E., Perez-Sanagustin, M., Ruiz-Calleja, A., Santos, P., Prieto, L., Serrano-Iglesias, S., & Väljataga, T. (2020). Learning design and learning analytics in mobile and ubiquitous learning: A systematic review. *British Journal of Educational Technology, 51* (4), 1078–1100.

    To further develop the reader's understanding of learning design and learning analytics in mobile and ubiquitous learning, we recommend this paper which offers an overview and analysis of current research. It proposes addressing mobile and ubiquitous learning beyond higher education settings, reinforcing the link between physical and virtual learning spaces, and more systematically aligning learning design and learning analytics processes.